# Wave scattering by *PT*-symmetric epsilon-near-zero periodic structures

O.V. Shramkova, G.P. Tsironis
Crete Center for Quantum Complexity and Nanotechnology, Department of Physics
University of Crete
Heraklion, Greece
oksana@physics.uoc.gr, gts@physics.uoc.gr

*Abstract*—The optical properties of *PT*-symmetric epsilon-near-zero (ENZ) periodic stack of the layers with balanced loss and gain have been examined. The effect of periodicity on the unidirectional tunneling phenomenon and symmetry breaking is determined. The performed analysis provides insight in the main features of the second harmonic (SH) generation by the ENZ *PT*-symmetric periodic structure.

## I. Introduction

The tremendous progress in the field of parity and time-reversal (*PT*-) symmetric quantum theory has been made during the last decade. *PT*-symmetric systems are non-Hermitian and invariant under the combined action of a parity- and time-reversal operation. The *PT*-related concepts can be realized in artificial optical materials that rely on balanced gain and loss regions. In this framework, *PT*-symmetry demands that the complex refractive index obeys the condition $n(\vec{r}) = n^*(-\vec{r})$. It was demonstrated that *PT*-symmetric materials can exhibit several exotic features, including unidirectional invisibility, coherent perfect absorption and nonreciprocity of light propagation [1]-[5].

Recently, the metamaterials with ENZ permittivities have attracted great attention owing to the specified wave interaction properties. It was shown that working in the ENZ regime with vanishingly small real part of dielectric permittivity we can obtain a markedly visible tunneling phenomenon at low levels of gain [6] in *PT*-symmetric bilayer. The aim of this paper is to explore the scattering properties of the linear and nonlinear *PT*-symmetric binary periodic ENZ material with balanced loss and gain.

## II. Problem Statement and Methodology

Let us consider the periodic structure composed of stacked isotropic homogeneous layers with relative dielectric permittivities $\varepsilon = \varepsilon' - i\varepsilon''$ and $\varepsilon^* = \varepsilon' + i\varepsilon''$ ($\varepsilon'$ and $\varepsilon''$ are positive) corresponding to balanced gain and loss regions. The system is *PT*-symmetric about $z=0$. The layers have identical thicknesses $d$ and are assumed of infinite extent in the *x-y*-plane. The stack is surrounded by linear homogeneous medium with dielectric permittivity $\varepsilon_a$ at $z \leq -L/2$ and $z \geq L/2$ as shown in Fig. 1, where $L=2Nd$ ($N$ is the number of structure periods) is the total thickness of the structure. The stack is illuminated by the plane wave incident at angle $\theta_i$. Since the layers are assumed isotropic in the *x-y* plane, the TE and TM polarised waves with the fields independent of the *y*-coordinate ($\partial/\partial y = 0$) can be analysed separately. Only the TM waves with the field components $E_x, E_z, H_y$ are discussed in the rest of the paper.

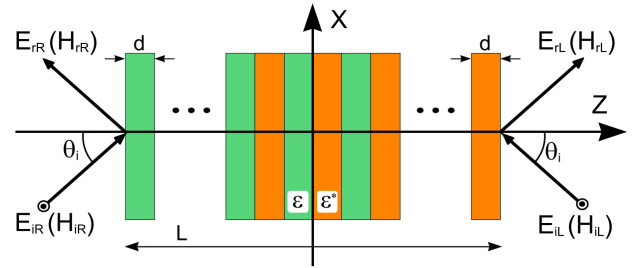

Fig. 1. Geometry of the problem

Let us study the *PT*-symmetric system from the point of view of the scattering matrix (*S*-matrix). The magnetic field of TM wave in the dielectric layer and surrounding homogeneous medium can be presented in the following form

$$H_y(\omega,x,z) = e^{-i\omega t + ik_x x} \begin{cases} Ae^{ik_{za}(z+L/2)} + Be^{-ik_{za}(z+L/2)}, & z \leq -L/2 \\ Ce^{ik_{za}(z-L/2)} + De^{-ik_{za}(z-L/2)}, & z \geq L/2 \end{cases}, \quad (1)$$

where $k_{za} = k_0\sqrt{\varepsilon_a}\cos\theta_i$ is the longitudinal wave number outside the stack; *A*, *B*, *C* and *D* are amplitudes of the forward and backward propagating waves outside the stack. These amplitudes are readily obtained by satisfying the continuity conditions for the tangential field components at the stack interfaces. This yields in the following relation

$$\begin{pmatrix} A \\ B \end{pmatrix} = \frac{1}{2} \begin{pmatrix} 1 & \frac{\omega}{c} \frac{\varepsilon_a}{k_{za}(\omega)} \\ 1 & -\frac{\omega}{c} \frac{\varepsilon_a}{k_{za}(\omega)} \end{pmatrix} \hat{M}(\omega) \begin{pmatrix} 1 & 1 \\ -\frac{c}{\omega} \frac{k_{za}(\omega)}{\varepsilon_a} & \frac{c}{\omega} \frac{k_{za}(\omega)}{\varepsilon_a} \end{pmatrix} \begin{pmatrix} D \\ C \end{pmatrix},$$

(2)

where $\hat{M}(\omega) = (\hat{m}(\omega))^N$ is the transfer matrix of the whole stack that can be expressed in terms of the transfer matrix $\hat{m}(\omega) = \hat{m}_{L1}(\omega)\hat{m}_{L2}(\omega)$ of a single period using Abeles theorem [7], where $\hat{m}_{L1,L2}(\omega)$ are the transfer matrices of each layer. Making use of (2) we can deduce the *S*-matrix in the form of the stack transfer matrix elements. The *S*-matrix is defined by

$$\begin{pmatrix} B \\ C \end{pmatrix} = \hat{S} \begin{pmatrix} A \\ D \end{pmatrix} = \begin{pmatrix} R^{(L)}(\omega) & T(\omega) \\ T(\omega) & R^{(R)}(\omega) \end{pmatrix} \begin{pmatrix} A \\ D \end{pmatrix}. \quad (3)$$

Here $R^{(L,R)}(\omega)$ are stack reflection coefficients for wave incident from the left and right, $T(\omega)$ is the transmission coefficient (it is the same for left and right incidence). The scattering matrix in this definition measures the breaking of *PT* symmetry. An alternative definition of the scattering matrix can be used for the determination of unidirectional invisibility points and takes the form

$$\begin{pmatrix} D \\ A \end{pmatrix} = \hat{S}_c \begin{pmatrix} B \\ C \end{pmatrix} = \begin{pmatrix} T(\omega) & R^{(L)}(\omega) \\ R^{(R)}(\omega) & T(\omega) \end{pmatrix} \begin{pmatrix} B \\ C \end{pmatrix}. \quad (4)$$

### III. TUNNELING CONDITIONS

As it was shown in [6], the tunneling phenomenon in ENZ bilayers with balanced loss and gain is mediated by the excitation of a surface wave at the interface separating the gain and loss regions. Authors of the mentioned work have determined the tunneling conditions and analysed three regimes of operation connected with gain/loss level.

In our work the analytical study of expressions for reflection coefficient $R^{(L,R)}(\omega)$ has revealed that for *PT*-symmetric ENZ periodic stack the tunnelling/zero-reflection conditions can be fulfilled if for sufficiently thick layers the angle of wave incidence approaches the critical angle

$$\theta_c = \arcsin\sqrt{\frac{\varepsilon'^2 + \varepsilon''^2}{2\varepsilon'}}. \quad (5)$$

This expression for the critical angle coincides with the expression obtained for *PT*-symmetric ENZ bilayers in [6]. Moreover, similar to the case for two slabs the plane wave incident at $\theta_i > \theta_c$ can excite the surface waves at the gain-loss interfaces (Fig.2). The peaks of distribution at the interfaces between layers in the period confirm this fact. All regimes of operation considered for *PT*-symmetric ENZ bilayers are valid for binary periodic *PT*-symmetric stacks.

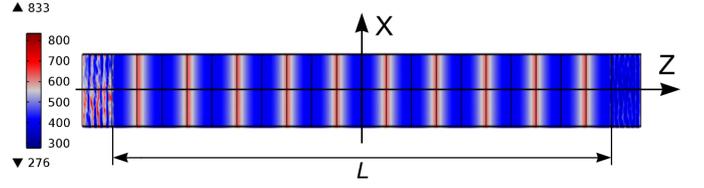

Fig. 2. Distribution of the magnetic-field magnitude $|H_y|$ for incidence from left at $\theta_i = 5°$, $\omega = 2.947 \times 10^{13}\ s^{-1}$, $\varepsilon' = 0.0001$, $\varepsilon'' = 0.001$, $d=125$ μm, $N=10$ (The numerical simulation was carried out by software package COMSOL Multiphysics).

### IV. *PT*-SYMMETRY BREAKING

Note that eigenvalues of the *S*-matrix in the *PT*-symmetric phase are unimodular and they must have reciprocal moduli or condition $|\lambda_1 \lambda_2| = 1$ should be satisfied ($\lambda_{1,2}$ are the eigenvalues of the *S*-matrix of a *PT*-symmetric layered structure). As a result we get that for symmetric phases $|\lambda_1| = |\lambda_2| = 1$. As it was demonstrated in the literature [8], a 1D *PT*-symmetric structures can undergo spontaneous symmetry-breaking transitions in the eigenvalues and eigenvectors of its *S*-matrix when an abrupt phase transition to a complex eigenspectrum takes place. In the *PT*-symmetry breaking points both eigenvalues meet and bifurcate, and the broken phases correspond to $|\lambda_1| = 1/|\lambda_2| > 1$. The anisotropic transmission resonance (ATR) [1],[8] for which *T*=1 and one of the reflection coefficients vanishes, can be observed in *PT*-symmetric periodic structures. This phenomenon can be referred to as unidirectional reflectivity and is associated with the exceptional points for the eigenvalues $\lambda_{1c,2c}$ of the $S_c$-matrix corresponding to the tunnelling conditions for left and right incidence.

We have demonstrated that the number *N* of stacked unit cells and thickness of the whole stack have strong impact on the transition and ATR. The numerical examination of frequency dependences for $\lambda_{1,2}$ in Figs.3(a)-3(b) shows that with increase of *N* (increase of total thickness of the stack) symmetry breaking occurs at lower frequency. In Fig.3(b) for *N*=100 the transition points reenter the symmetric phase at the frequencies close to Bragg resonances. The numerical analysis of the symmetry breaking transition for 2 different ENZ-stacks with the same overall stack thicknesses *L*=250 μm but increasing number of the periods and thinner constituent layers is presented in Fig.3(a) and Fig.3(c). It can be observed that, for increased number of unit cells in the stack the transition tends to occur at higher frequencies. The analytical study has revealed that the fine-stratified periodic structure with layer thicknesses much less than the wavelength of electromagnetic wave can be considered as anisotropic dielectric with real effective components of the tensor of dielectric permittivity ($\varepsilon_{xx} = \varepsilon'$ and $\varepsilon_{zz} = \dfrac{\varepsilon'^2 + \varepsilon''^2}{\varepsilon'}$). So, the symmetry-breaking transitions will not be observed in fine-stratified *PT*-symmetric structures.

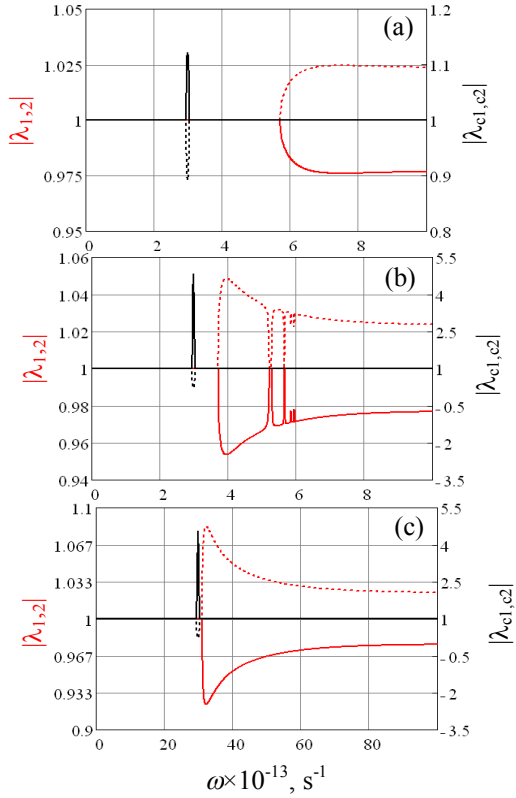

Fig. 3. Modulus of eigenvalues of *S*- and $S_c$- matrixes for 1D *PT*-symmetric periodic stack of the layers as a function of frequency at $\varepsilon' = 0.0001$, $\varepsilon'' = 0.001$, $\theta_i = 5°$; (a) - $d$=125 μm, *N*=1; (b) - $d$=125 μm, *N*=100; (c) - $d$=12.5 μm, *N*=10. Red curves correspond to the eigenvalues $\lambda_{1,2}$, black curves are for $\lambda_{1c,2c}$; solid and dashed curves of the same color correspond to the different eigenvalues of the same scattering matrix.

The exceptional points for $S_c$-matrix do not depend on the number of the periods (Figs.3(a)-3(b)). At the same time it can be seen that the phenomenon of unidirectional reflectivity is essentially dependent on the thickness of the constituent binary layers (Figs.3(a)-3(c)). This means that the frequencies of ATR are determined only by the materials and thickness of the layers.

The numerical analysis of reflectivity/transmittivity for ENZ *PT*-symmetric periodic stack (parameters as in Fig.3(b)) is presented in Fig. 4. Two very closely spaced ATRs (one for left incidence and one for right) correspond to frequencies $\omega = 2.947 \times 10^{13}$ $s^{-1}$ and $\omega = 3.054 \times 10^{13}$ $s^{-1}$. As it was mentioned before these frequencies do not depend on the number of structure periods. At the same time the magnitudes of reflectivity/transmittivity will be changed close to the points of ATR. All other resonances (Bragg resonances) occur as the stack overall thickness equals an integer number of Bloch half-waves. The number of points corresponding to Bragg resonances depends on the number of periods *N*. The amplification of reflected waves is connected with a constructive interaction between the forward- and backward-propagating waves.

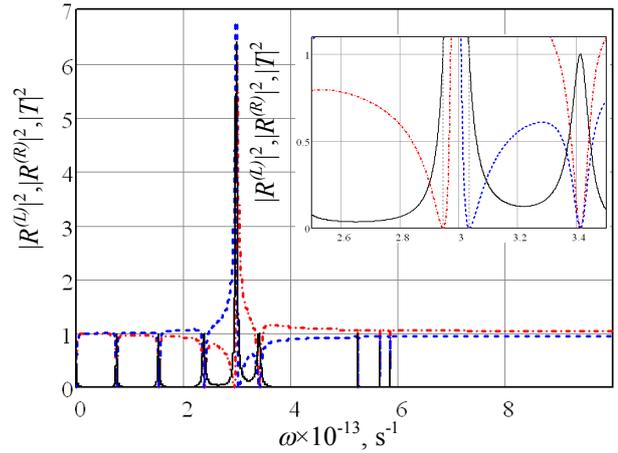

Fig. 4. Reflectance ($|R^{(L)}|^2$ – red dash-dot curve, $|R^{(R)}|^2$ – blue dashed curve) and transmittance (black solid curve) for *PT*- symmetric periodic layered structure versus frequency of plane TM wave incident at $\theta_i = 5°$; $\varepsilon' = 0.0001$, $\varepsilon'' = 0.001$, $d$=125 μm, *N*=100. Inset: zoomed in onto the 2 ATRs.

## V. SECOND HARMONIC GENERATION

In this section, the distinctive features of the TM wave scattering by the nonlinear *PT*-symmetric periodic stack, illuminated by a pump wave of frequency $\omega$, are illustrated by the results of detailed parametric study.

Let us assume that the layers of the periodic structure are characterised by the second-order nonlinear susceptibility $\chi^{(2)}$. The scattering characteristics of the TM waves at the SH frequency $\omega_{SH}=2\omega$, generated through the three-wave mixing process inside the layers [9], are obtained in the approximation of weak nonlinearity. The full solution of inhomogeneous Helmholtz equation can be represented as a superposition of the 6 partial plane waves which must satisfy the conditions of field continuity at the layer interfaces. Also, the requirement of the waveform invariance along the layer interfaces imposes the additional constraint on the *x*-components of the wave vectors of the interacting waves, which must obey the phase synchronism condition in the three-wave mixing process: $k_{xSH} = 2k_x$, where $k_x = k_0\sqrt{\varepsilon_a}\sin\theta_i$. Enforcing the boundary conditions of field continuity at the interfaces of the stacked layers and using the modified transfer matrix method[10], we obtain the nonlinear scattering coefficients $F_r$ at $z < -L/2$ and $F_t$ at $z > L/2$ describing the field of frequency $\omega_{SH}$ emitted from the stack into the surrounding homogeneous medium.

The numerical analysis of the SH generation process by the nonlinear layered *PT*-symmetric structure illuminated by a pump waves of frequency $\omega$ incident at angle $\theta_i$ has been performed at the following parameters of the nonlinear layers: $\varepsilon' = 0.0001$, $\varepsilon'' = 0.001$, $d = 125$ μm, $\chi^{(2)} = 2 \times 10^{-7}$ m/V.

The field intensities $|F_{r,t}|$ of the SH frequency $\omega_{SH}$ emitted from the stack generated in the structure with $N=10$ are shown in Fig. 5 for variable frequency $\omega$ close to ATRs. These dependencies demonstrate correlation between $|F_{r,t}|$ and $|T(\omega)|$. The pump wave amplification has significant effect on the SH frequency. It is evident here that the efficiency of SH generation is hardly improved at the frequency between 2 ATRs for which we obtain peaks of reflectivity and transmittivity.

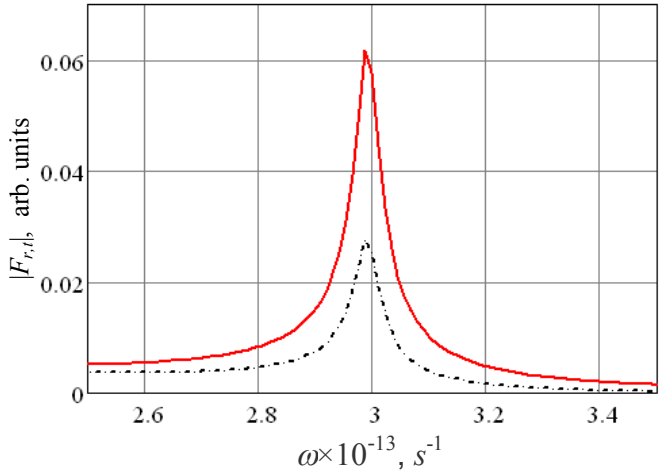

Fig. 5. The intensity of field radiated at SH frequency $\omega_{SH}$ in the reverse (solid red line) and forward (dash-dot black line) directions of the $z$-axis from periodic stack, $N=10$. The stack is illuminated by pump wave incident at $\theta_i = 5°$.

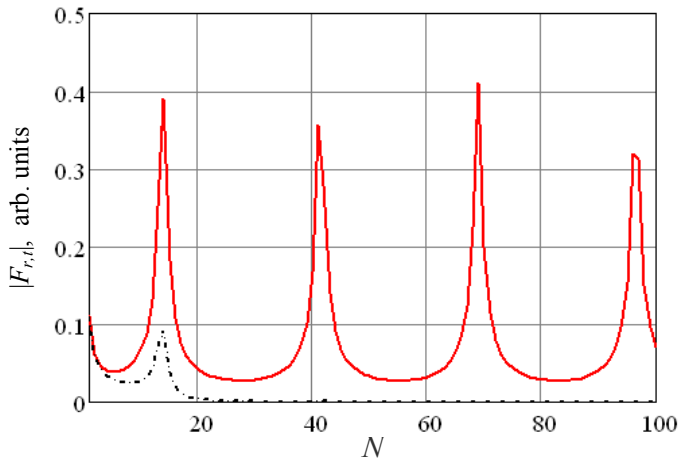

Fig. 6. The intensity of field radiated at SH frequency $\omega_{SH}$ in the reverse (solid red line) and forward (dashed black line) directions of the $z$-axis from periodic stack. The stack is illuminated by pump wave of frequency $\omega = 2.988 \times 10^{13}$ $s^{-1}$ incident at $\theta_i = 5°$.

The nonmonotonic dependences of $|F_{r,t}|$ on the number $N$ of unit cells in the stack are presented in Fig.6. For the presented simulations the chosen pump wave frequency corresponds to the frequency for peak intensity in Fig.5. Fig.6 shows that $|F_r|$ has maxima at $N = 14, 41, 69, 96$, whereas $|F_t|$ has a higher peak at $N = 14$ and then falls with $N$. It is necessary to note that the $|F_r|$ maxima are well correlated with the transmittance maxima of pump wave.

It was obtained that intensities of scattered SH waves are the same for left and right incidence. Moreover, for $PT$-symmetric structures with equal total thicknesses but different number of periods and thicknesses of the layers the nonlinear response will be the same.

## VI. CONCLUSION

The spectral features of $PT$-symmetric ENZ periodic structure have been explored. The effect of the structure geometry on the tunnelling phenomenon and spontaneous symmetry breaking is demonstrated.

The properties of the SH generation by $PT$-symmetric periodic stacks of binary ENZ nonlinear layers have been analysed. The effects of the structure parameters and the incident pump wave characteristics on the efficiency of the SH generation have been investigated in detail. The strong enhancement of the SH generation efficiency at the resonant point between 2 ATRs is obtained.

ACKNOWLEDGMENT

The research work was partially supported by the European Union Seventh Framework Program (FP7-REGPOT-2012-2013-1) under grant agreement No. 316165